\documentclass[9pt,twocolumn,twoside]{opticajnl}

\journal{opticajournal} 

\setboolean{shortarticle}{true}

\usepackage{lineno}

\newcommand{\eps}{\varepsilon}

\newcommand{\vc}[1]{{\bf #1}}

\renewcommand{\c}{\chi}

\begin{document}

\title{Multipole generalization of the Witten effect in Mie-resonant photonics}

\author[1,*]{Timur Seidov}
\author[1]{Maxim A. Gorlach}

\affil[1]{School of Physics and Engineering, ITMO University, Saint-Petersburg 197101, Russia}
\affil[*]{tim.phys.math@gmail.com}

\begin{abstract}
We present a generalization of the Witten effect on the case of oscillating multipole sources exciting nonreciprocal sphere with effective axion response. We find that the fields outside of the sphere are presented as a superposition of electric and magnetic multipoles. In addition to appearance of cross-polarized component in the radiation, Mie resonances of the system hybridize with each other, exhibiting characteristic double peaks in Mie spectra observed especially clearly for higher-order multipole resonances. This characteristic feature may provide a sensitive probe of axion-type nonreciprocal responses in Mie-resonant photonics.
\end{abstract}

\maketitle


The equations of axion electrodynamics were originally introduced in high energy physics to describe the coupling of the hypothetic axion~\cite{Peccei_Axion_Origin,Wilczek_Axion_Origins,Weinberg_Axion_Origin} to the electromagnetic field. While axions are yet to be observed, axion field arises as emergent property in a range of condensed matter and photonic systems~\cite{Wilczek_Applications,Nenno_Axion_Condensed,Sekine2021} giving rise to rich physics. 

In condensed matter context, the relevant systems are antiferromagnets~\cite{Dzyaloshinskii1960,Astrov1960}, magnetoelectrics~\cite{Nenno_Axion_Condensed} and topological insulators~\cite{Sekine2021} with relatively weak axion response. In photonics, a similar concept has recently drawn much attention because of the prospect to tailor and enhance effective axion fields~\cite{Prudencio_Space-Time_Crystals,Shaposhnikov_Emergent_Axion,SafaeiJazi2024,Seidov2024Tellegen}. Recent experiments~\cite{Yang2025,BaileZhang2025} have confirmed that expectation opening further avenues for nonreciprocal photonics.




One of the earliest predictions of axion electrodynamics is the Witten effect~-- emergence of effective dyon charges from the magnetic monopole placed in the axion field~\cite{Witten}. This effect arises in condensed matter as well~\cite{Rosenberg_Witten_effect}, while its modifications include static magnetic dipole in the oscillating axion field~\cite{Hill_Dipole_Moment} or oscillating magnetic dipole in the static axion field~\cite{Seidov2023}. In both cases, the system acquires an effective electric dipole moment. Such hybridization of electric and magnetic dipole modes provides a signature of $\mathcal{T}$-breaking phenomena allowing to distinguish various nonreciprocal responses~\cite{Seidov2024}.


However, dipole resonances possess relatively low quality factors, which limits the capacity to probe material nonreciprocity. In contrast, higher-order multipole resonances are much narrower, potentially providing a more sensitive probe. This motivates us to explore a multipole generalization of the Witten effect, when electric or magnetic multipole source excites a Mie-resonant particle with permittivity $\eps$, permeability $\mu$ and time-independent effective axion response $\chi$.




Axion electrodynamics is a modification of Maxwell's equations generated by the additional term $L_\theta = \chi/(4\pi\,c)\,(\mathbf{E}\cdot \mathbf{B})$ introduced in the Lagrangian. The same modification of Maxwell's equation is obtained in the medium with a special type of constitutive relations
\begin{equation}
\mathbf{D} = \varepsilon\,\mathbf{E} + \chi\,\mathbf{B}\, , 
\quad
\mathbf{H} = \mu^{-1} \, \mathbf{B} - \chi\,\mathbf{E} \, ,
\end{equation}
where $\chi$ measures the strength of the axion response. In photonics, such media are called \textit{Tellegen} media, while real-valued $\chi$ is often termed Tellegen parameter. In the case of a homogeneous time-independent Tellegen parameter, its presence is manifested only at the boundary. This leads to a special type of boundary conditions:
\begin{gather}
B_n^{1} = B_n^{2}, 
\quad
\mu_{1}^{-1} \vc B_\tau^{1} - \mu_{2}^{-1} \vc B_\tau^{2} = \chi \, \vc E_\tau,
\label{eq:BC_1} \\
\vc E_\tau^{1} = \vc E_\tau^{2}, 
\quad
\varepsilon_{1}\, E_n^{1} - \varepsilon_{2}\, E_n^{2} = -\,\chi\, B_n.
\label{eq:BC_2}
\end{gather}

In our work, we consider a sphere of radius $R$ with Tellegen parameter $\chi$, permittivity $\eps$ and permeability $\mu$ excited by electric or magnetic multipole source $Q^s_{lm}$ placed in its center. Below, we calculate the structure of the fields inside and outside identifying characteristic features due to the nonzero axion field $\chi$.


To solve the problem, we use multipole expansion method and the basis of vector spherical harmonics~\cite{jackson}. The fields inside take form
\begin{gather}
\vc E_I \;=\; \frac{1}{\sqrt{\eps}}\sum_{l,m} \Bigl( 
\frac{i}{\tilde q} \nabla \times \left[\left( a_{e,\, s}^{l,m} \, h_l^{(1)}(\tilde qr) + a_{e,\, i}^{l,m} \, j_l (\tilde qr)\right) \, \vc X_{l,m} \right]
\;-\; \nonumber \\
 a_{m,\, i}^{l,m} \, j_l(\tilde qr) \, \vc X_{l,m}
\Bigr),
\label{eq:expansion_E_I}
\end{gather}
\begin{gather}
\vc H_I \;=\; \frac{1}{\sqrt{\mu}}\sum_{l,m} \Bigl( 
\frac{i}{\tilde q} \nabla \times \left [ a_{m,\, i}^{l,m} \, j_l(\tilde qr) \, \vc X_{l,m} \right] 
\;+\; \nonumber \\
\left( a_{e,\, s}^{l,m} \, h_l^{(1)}(\tilde qr) + a_{e,\, i}^{l,m} \, j_l (\tilde qr)\right) \, \vc X_{l,m}
\Bigr),
\label{eq:expansion_B_I}
\end{gather}
where spherical Hankel function of the first kind $h_l^{(1)}$  corresponds to the outgoing wave radiated by point electric multipole, spherical Bessel functions $j_l$ captures the contribution of the reflected field, $\tilde q=\sqrt{\eps\mu}\,\omega/c$ is a wave number, while $\vc X_{l,m}$ is a vector spherical harmonic with indices $(l,m)$. Coefficients $a_{e,\, s}^{l,m}$ are known multipole expansion coefficients of the sources, while $a_{e,\, i}^{l,m}$ and $a_{m,\, i}^{l,m}$ are unknown. Fields outside of the sphere contain only outgoing waves and thus read
\begin{gather}
\vc E_O \;=\; \sum_{l,m} \Bigl( 
\frac{i}{q} \nabla \times \left[ a_{e}^{l,m} \, h_l^{(1)}(qr) \, \vc X_{l,m} \right]
\;-\; \nonumber \\
 a_m^{l,m} h_l^{(1)}(qr) \, \vc X_{l,m}
\Bigr),
\label{eq:expansion_E_O} \\
\vc H_O \;=\; \sum_{l,m} \Bigl( 
\frac{i}{q} \nabla \times \left [ a_m^{l,m} h_l^{(1)}(qr) \, \vc X_{l,m} \right] 
\;+\; \nonumber \\
  a_{e}^{l,m} \, h_l^{(1)}(qr)  \, \vc X_{l,m}
\Bigr),
\label{eq:expansion_B_O}
\end{gather}
where $q=\omega/c$. By choosing surface normal vector as the radial one and substituting the expansions \eqref{eq:expansion_E_I}-\eqref{eq:expansion_B_O} into the conditions \eqref{eq:BC_1}, \eqref{eq:BC_2} for tangential components, we obtain four sets of linear equations. The conditions for the normal components of the fields are fulfilled automatically. For notation simplicity we introduce the operator
\begin{equation}
    D_{R}f(qR) \;=\; \frac{d}{dR}\,\Bigl(R\,f(qR)\Bigr)\, ,
\end{equation}
 and recall the property of vector spherical harmonics
 \begin{gather}
    \nabla\times[f(qr)\mathbf{X}_{l,m}]=\frac{1}{r}\underbrace{\frac{d}{dr}[r f(qr)]}_{D_r f(qr)} \underbrace{\mathbf{n}\times\mathbf{X}_{l,m}}_{\mathbf{Z}_{l,m}}+ \nonumber \\ i\sqrt{l(l+1)}\frac{f(qr)}{r}Y_{l,m}\mathbf{n}. \label{VSHprop}
\end{gather}

\begin{figure*}[ht]
    \centering
    \includegraphics[width=0.85\textwidth]{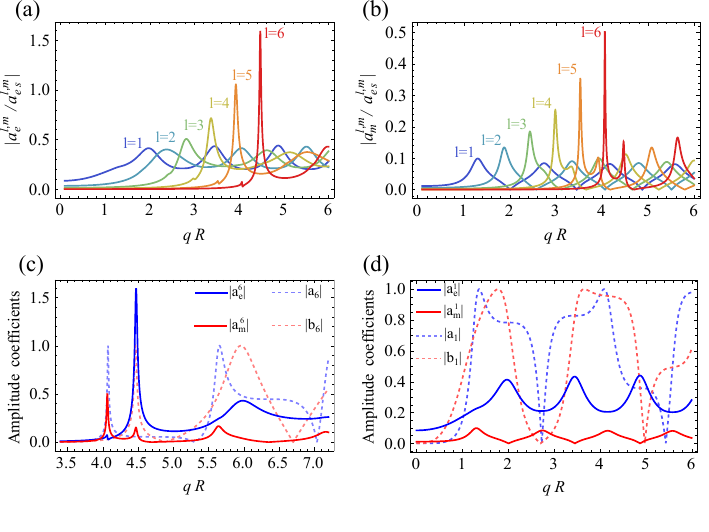}
    \caption{(a,b) Electric (a) and magnetic (b) multipoles of the field outside the axion sphere normalized to the respective electric multipole moment of the source versus frequency. Different colors correspond to the different order of the multipoles.
    (c,d) Electric $a^l_e$ and magnetic $a^l_m$ multipoles outside the sphere normalized to the multipole moment of the source compared to the standard Mie coefficients $a_l$ \eqref{eq:Mie_al} and $b_l$ \eqref{eq:Mie_bl} for $l=6$ (c) and $l=1$ (d). In all calculations $\eps=5,\mu=1,\chi=0.4$.}
    \label{fig:Fig1}
\end{figure*}

Thus, we arrive at the following system

\begin{gather}
    a_{m}^{l,m} h^{(1)}_l(q)=\frac{a_{m \, i}^{l,m}}{\sqrt{\eps}}j_l(\tilde q)\, ,\\
    a_{e}^{l,m}=\frac{1}{\varepsilon\sqrt{\mu}}\Bigg(a_{e,\,s}^{l,m}\frac{D_R h^{(1)}_l(\tilde q)}{D_R h^{(1)}_l(q)}+a_{e,\,i}^{l,m}\frac{D_R j_l(\tilde q)}{D_R h^{(1)}_l(q)}\Bigg)\, , \\
    \frac{1}{\sqrt{\mu}}(a_{e,\,s}^{l,m} h^{(1)}_l(\tilde q) + a_{e,\,i}^{l,m} j_l(\tilde q))- a_{e}^{l,m} h^{(1)}_l(q)= -\chi \frac{a_{m \, i}^{l,m}}{\sqrt{\eps}}j_l(\tilde q)\, , \\
    \begin{split}
            \frac{a_{m \, i}^{l,m}}{\mu\sqrt{\varepsilon}} D_R j_l(\tilde q) -  a_{m}^{l,m} D_R h^{(1)}_l(q)= \\ \frac{\chi}{\varepsilon\sqrt{\mu}} \Bigg(a_{e,\,s}^{l,m}D_R h^{(1)}_l(\tilde q)+a_{e,\,i}^{l,m}D_R j_l(\tilde q)\Bigg)\:.\label{eq:med_sys4}
    \end{split}
\end{gather}
The solution of this system is expressed using auxiliary coefficients
\begin{gather}
    \alpha_l=\frac{j_l(\tilde q R)}{h_l^{(1)}(q R)}, \quad
    \beta_l=\frac{D_R j_l(\tilde q R)}{D_R h_l^{(1)}(q R)}, \\
    \gamma_l=\frac{h_l^{(1)}(\tilde q R)}{h_l^{(1)}(q R)}, \quad
    \delta_l=\frac{D_R h_l^{(1)}(\tilde q R)}{D_R h_l^{(1)}(q R)}
\end{gather}
and reads
\begin{gather}
    a_{e}^{l,m}=\frac{a_{e,\,s}^{l,m}}{\sqrt{\mu}}\frac{(\alpha_l\mu-\beta_l)(\alpha_l\delta_l-\beta_l\gamma_l)}{\alpha_l\eps(\alpha_l \mu-\beta_l)+\beta_l(\beta_l-\alpha_l\mu(\chi^2+1))}\, , \label{eq:sph_solve1} \\
    a_{m}^{l,m}= - a_{e,\,s}^{l,m}\frac{\chi\sqrt{\mu}\alpha_l(\alpha_l\delta_l-\beta_l\gamma_l)}{\alpha_l\eps(\alpha_l \mu-\beta_l)+\beta_l(\beta_l-\alpha_l\mu(\chi^2+1))}\, , \label{eq:sph_solve2}\\
    a_{e,\,i}^{l,m}=a_{e,\,s}^{l,m}\frac{\eps\gamma_l(\beta_l-\mu\alpha_l)-\beta_l\delta_l+\mu\alpha_l\delta_l(\chi^2+1)}{\alpha_l\eps(\alpha_l \mu-\beta_l)+\beta_l(\beta_l-\alpha_l\mu(\chi^2+1))}\, ,\\
    a_{m,\,i}^{l,m}=-a_{e,\,s}^{l,m}\frac{\chi \sqrt{\eps\mu}(\alpha_l\delta_l-\beta_l\gamma_l)}{\alpha_l\eps(\alpha_l \mu-\beta_l)+\beta_l(\beta_l-\alpha_l\mu(\chi^2+1))}\, . \label{eq:sph_solve4}
\end{gather}

Multipole expansion coefficients outside of the sphere are related to the multipole moments as~\cite{jackson}
\begin{gather}
    a_{e}^{l,m}=\frac{cq^{l+2}}{i(2l+1)!!}\sqrt{\frac{l+1}{l}}Q_{lm}^{\text{eff}}\, , \label{a_e} \\
    a_{m}^{l,m}=\frac{iq^{l+2}}{(2l+1)!!}\sqrt{\frac{l+1}{l}} M_{lm}^{\text{eff}}\, . \label{a_m}
\end{gather}
The same is true for the multipoles inside the sphere with the additional prefactors $1/\sqrt{\eps}$ and $1/\sqrt{\mu}$. Thus, the ratio between the multipole moments of the source and the respective multipole moments of the field outside reads
\begin{gather}
    \frac{Q_{lm}^{\text{eff}}}{Q_{lm}^{s}}=\underbrace{\frac{a_{e}^{l,m}}{a_{e, \, s}^{l,m}}}_{a^{l}_e}\eps^{\frac{l+1}{2}}\mu^{\frac{l+2}{2}} \, , \quad
    \frac{M_{lm}^{\text{eff}}}{Q_{lm}^{s}}=-c\underbrace{\frac{a_{m}^{l,m}}{a_{e, \, s}^{l,m}}}_{a^{l}_m} \eps^{\frac{l}{2}}\mu^{\frac{l+1}{2}} \, .
\end{gather}
Note that even though the sphere is excited by electric multipole source, the field outside contains magnetic multipole contribution. This provides a signature of $\mathcal{T}$ symmetry breaking in the material and a multipole generalization of the Witten effect.

As anticipated from the rotational symmetry of the system and seen from \eqref{eq:sph_solve1}-\eqref{eq:sph_solve4}, the ratios of the multipole moments do not depend on $m$. Therefore, we set $m=0$ and plot these ratios for various $l$ versus source frequency in Fig.\ref{fig:Fig1}.

As in the dipole case, we observe that each of effective magnetic multipoles of the structure vanishes at several specific frequencies. We argue that this is due to the boundary conditions, as there are surface currents flowing at the boundary of the sphere and leading to the cancellation of certain magnetic multipoles. Thus, we interpret such zeros of ratios as a multipole generalization of magnetic anapole states~\cite{Mirosh2015} of the axion sphere. It is straightforward to verity that such states are a property of the sphere itself by solving the corresponding eigenmode problem. 

However, differently from the dipole case, hybridization of electric and magnetic multipoles not only adds a cross-polarized component to the far-field radiation, but also is clearly manifested in the amplitude ratio coefficients (Fig.~\ref{fig:Fig1}), separately for co-polarized (a) and cross-polarized components (b), especially for the resonances with high quality factor. This becomes possible because effective axion field $\chi$ hybridizes the Mie modes of the particle corresponding to pure electric or magnetic multipoles. Such hybridization is barely noticeable in the dipole case due to the significant width of the resonances, but becomes more pronounced for higher-order multipoles. Note that Mie coefficients $a_6,b_6$ are calculated using classical formulae~\cite{jackson}
\begin{gather}
    a_l \;=\;
    \frac{
    \sqrt{\eps\mu}\,\psi_l\bigl(\sqrt{\eps\mu} q R\bigr)\,\psi_l(q R)
    \;-\;
    \psi_l'(q R)\,\psi_l\bigl(\sqrt{\eps\mu} q R\bigr)
    }{
    \sqrt{\eps\mu}\,\psi_l'\bigl(\sqrt{\eps\mu} q R\bigr)\,\xi_l(q R)
    \;-\;
    \psi_l\bigl(\sqrt{\eps\mu} q R\bigr)\,\xi_l'(q R)}
    , \label{eq:Mie_al} \\
    b_l \;=\;
    \frac{
    \psi_l'\bigl(\sqrt{\eps\mu} q R\bigr)\,\psi_l(q R)
    \;-\;
    \sqrt{\eps\mu}\,\psi_l'(q R)\,\psi_l\bigl(\sqrt{\eps\mu} q R\bigr)
    }{
    \psi_l'\bigl(\sqrt{\eps\mu} q R\bigr)\,\xi_l(q R)
    \;-\;
    \sqrt{\eps\mu}\,\xi_l'(q R)\,\psi_l\bigl(\sqrt{\eps\mu} q R\bigr)}
    , \label{eq:Mie_bl}  \\
    \psi_l(z) \;=\; z\,j_l(z), 
    \quad
    \xi_l(z) \;=\; z\,h_l^{(1)}(z) \nonumber,
\end{gather}
and are merely used in Fig.~\ref{fig:Fig1}(c) to identify the frequencies of the Mie eigenmodes. These coefficients should not be compared directly to our amplitude ratios, since we calculate the problem with the sources inside, not the scattered-wave problem. Figure~\ref{fig:Fig1}(c) thus clearly represents hybridization of modes at the first pair of Mie resonances. Since further Mie resonances possess lower quality factors, as can be seen by the width of Mie coefficients' peaks, hybridization is less obvious after the first two resonances. In case of dipoles and lower multipoles, the hybridization is suppressed for the same reason, Fig.~\ref{fig:Fig1}(d).

Above, we focused on the excitation of the sphere by electric sources. For magnetic sources, the solution is qualitatively similar. We use the same fields \eqref{eq:expansion_E_O},(\ref{eq:expansion_B_O}) outside. However, as follows from \eqref{a_e},(\ref{a_m}), $a^{l,m}_{e,s}$ must be zero and instead $a^{l,m}_{m,s}$ must be added to the magnetic field. Thus, the fields inside take form 
\begin{gather}
\vc E_I \;=\; \frac{1}{\sqrt{\eps}}\sum_{l,m} \Bigl( 
\frac{i}{\tilde q} \nabla \times \left[a_{e,\, i}^{l,m} \, j_l (\tilde qr) \, \vc X_{l,m} \right]
\;-\; \nonumber \\
\left(a^{l,m}_{m,s} h_l^{(1)}(\tilde qr) + a_{m,\, i}^{l,m} \, j_l(\tilde qr)\right) \, \vc X_{l,m}
\Bigr),
\end{gather}
\begin{gather}
\vc H_I \;=\; \frac{1}{\sqrt{\mu}}\sum_{l,m} \Bigl( 
\frac{i}{\tilde q} \nabla \times \left [\left(a^{l,m}_{m,s} \, h_l^{(1)}(\tilde qr) + a_{m,\, i}^{l,m} \, j_l(\tilde qr)\right) \, \vc X_{l,m} \right] 
\;+\; \nonumber \\
a_{e,\, i}^{l,m} \, j_l (\tilde qr) \, \vc X_{l,m}
\Bigr),
\end{gather}
while the respective solution reads
\begin{gather}
a_{e,\,i}^{l,m}=a_{m,\,s}^{l,m} \frac{ \sqrt{\eps \mu} \chi (\alpha_l  \delta_l -\beta_l  \gamma_l )}{\alpha_l  \eps (\alpha_l  \mu-\beta_l )+\beta_l  \left(\beta_l -\alpha_l  \mu \left(\chi^2+1\right)\right)}
\, , \label{eq:sph_m_solve1} \\
a_{m,\,i}^{l,m}=a_{m,\,s}^{l,m} \frac{ \left(-\beta_l  \delta_l +\alpha_l  \eps (\delta_l -\gamma_l  \mu)+\beta_l  \gamma_l  \mu \left(\chi^2+1\right)\right)}{\alpha_l  \eps (\alpha_l  \mu-\beta_l )+\beta_l  \left(\beta_l -\alpha_l  \mu \left(\chi^2+1\right)\right)}
\, , \\
a_{e}^{l,m}=\frac{a_{m,\,s}^{l,m}}{\sqrt{\eps}}\frac{ \beta_l  \chi (\alpha_l  \delta_l -\beta_l  \gamma_l )}{ \left(\alpha_l  \eps (\alpha_l  \mu-\beta_l )+\beta_l  \left(\beta_l -\alpha_l  \mu \left(\chi^2+1\right)\right)\right)}\, ,  \label{eq:sph_m_solve3}  \\
a_{m}^{l,m}=\frac{a_{m,\,s}^{l,m}}{\sqrt{\eps}}\frac{ (\alpha_l  \eps-\beta_l ) (\alpha_l  \delta_l -\beta_l  \gamma_l )}{\left(\alpha_l  \eps (\alpha_l  \mu-\beta_l )+\beta_l  \left(\beta_l -\alpha_l  \mu \left(\chi^2+1\right)\right)\right)}\, .
\end{gather}
\begin{figure}
    \centering
    \includegraphics[width=\linewidth]{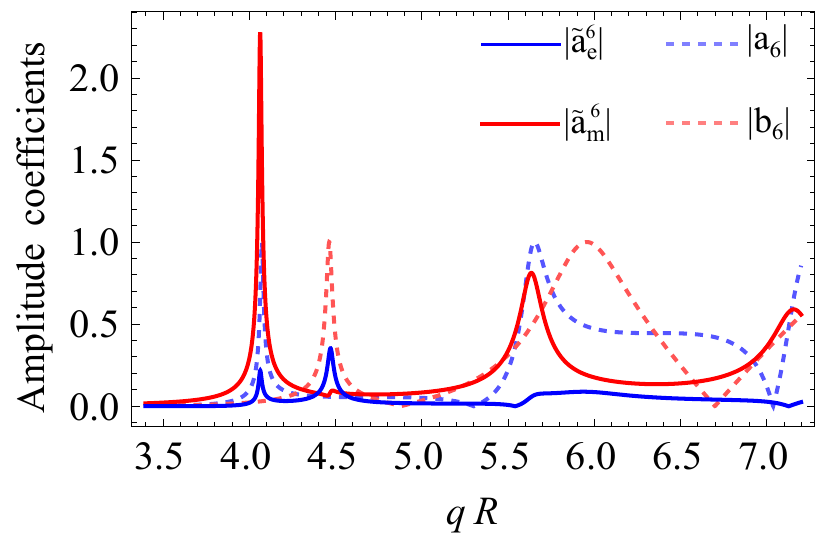}
    \caption{Electric $\tilde a^l_e$ and magnetic $\tilde a^l_m$ multipole moments of the field outside the axion sphere normalized to the respective magnetic multipole with $l=6$ of the source in the center of the sphere. Spectral features are compared to those of the Mie coefficients  $a_l$ \eqref{eq:Mie_al} and $b_l$ \eqref{eq:Mie_bl}. Parameters are $\eps=5,$ $\mu=1,\chi=0.4$.}
    \label{fig:Fig2}
\end{figure}
Figure~\ref{fig:Fig2} shows that the overall structure of the solution remains intact. However, given the presence of the magnetic source, magnetic resonances are significantly amplified, while electric resonances are weakened. This can be attributed to the fact that the hybridization of the multipoles is of the first order in the Tellegen parameter $\chi$, as is evident from \eqref{eq:sph_solve2} and \eqref{eq:sph_m_solve3}.


To conclude, we have provided a generalization of the Witten effect on the case of oscillating multipole sources in the static axion field which is a common situation in condensed matter and photonic context. For higher-order electric multipole sources, we have demonstrated not only the effective magnetic multipoles observed outside, but also the hybridization of electric and magnetic Mie resonances, featuring characteristic double peaks in Mie amplitude coefficients seen distinctly only for the high-quality higher-order multipole resonances.


This characteristic feature may be useful in experimental detection and distinction of axion-type responses as well as characterization of non-reciprocal meta-atoms~\cite{SafaeiJazi2024}, as such excitation scheme can readily be implemeted in experiments with microwave metamaterials~\cite{Yang2025} providing a sensitive probe of $\mathcal{P}$ and $\mathcal{T}$-breaking phenomena in nonreciprocal metamaterials and a useful platform for Mie-resonant nonreciprocal photonics.



Theoretical models were supported by Priority 2030 Federal Academic Leadership Program. Numerical simulations were supported by the Russian Science Foundation (Grant No.~23-72-10026). The authors acknowledge partial support from the Foundation for the Advancement of Theoretical Physics and Mathematics ``Basis''.

\bibliography{bibliography}

\end{document}